# Fast Crack Detection Using Convolutional Neural Network

Jiesheng Yang[1], Fangzheng Lin[1], Yusheng Xiang[2], Peter Katranuschkov[1], Raimar J. Scherer[1]
[1]Institute of Construction Informatics, Technische Universität Dresden, Dresden, Germany
[2]Department of Mechanical Engineering, Karlsruhe Institute of Technology, Karlsruhe, Germany
jiesheng.yang@tu-dresden.de

**Abstract.** To improve the efficiency and reduce the labour cost of the renovation process, this study presents a lightweight Convolutional Neural Network (CNN)-based architecture to extract crack-like features, such as cracks and joints. Moreover, Transfer Learning (TF) method was used to save training time while offering comparable prediction results. For three different objectives: 1) Detection of the concrete cracks; 2) Detection of natural stone cracks; 3) Differentiation between joints and cracks in natural stone; We built a natural stone dataset with joints and cracks information as complementary for the concrete benchmark dataset. As the results show, our model is demonstrated as an effective tool for industry use.

**Introduction**

In the field of non-destructive stone defect testing, different methods, e.g., Visual inspection Test (VT) Magnetic particle Testing (MT) and Ultrasonic Testing (UT), have been used to test surface defects. In practical use, every method has obvious imperfections: the MT equipment costs 904,48 € (UV Magnetic Yoke Flaw Detector - AJE-220 | Katex Ltd, 2021), and UT equipment costs 17.253,69 € (Olympus Panametrics Omniscan MX 32:128 Ultraschall Phased Feld Pa Flaw | eBay, 2021). Issues from high expense on MT equipment and UT equipment have not been adequately addressed. Moreover, disadvantages of VT can also not be overlooked. The test result of VT strongly depends on the tester's experience and thus subjective.

Hence, empowering VT with increasing computing power to improving efficiency and productivity and avoid the high cost of special equipment in renovation works is the starting point of this research. Abdel-Qader (2003) proved a Fast Haar Transform edge detection method in bridge crack identification. Based on that, Yeum and Dyke (2015) proposed a sliding windows technique in image processing techniques to detect cracks on steel. However, the results of edge detection are mainly affected by the noises. Thus, deep learning is employed in later application: Cha (2017) trained models to recognize concrete cracks with 97.95% accuracy on his test dataset. Similarly, Satyen (2019) achieved 85% accuracy with his test dataset in the crack recognition task. Unfortunately, those models are trained in labs with expensive powerful machines, for example, Cha performed all task on a workstation with two GPUs (CPU: Intel Xeon E5-2650 v3 @2.3GHz, RAM:64GB and GPU: Nvidia GeForce Titan X × 2ea), which limit the promotion of their approach.

As a consensus, stone crack images are more difficult to obtain compared with brick crack ones. At the same time, the TF can apply the weights of an already trained deep learning model to a different but related problem, and it shall be used if the old task has more data than the new task (Yosinski, 2015). Instead of starting the training process from scratch, this method starts with features that have been learned from a previous task, where a lot of labelled training data are available. As evidence, many studies in construction domain have mentioned that they have followed this idea and benefited from it (Xiang, 2020a; Xiang, 2020b). Hence the TF method is used to address this problem.

Given the above-mentioned facts, the scope of this research is focused on: Building a lightweight convolutional neural network (CNN) architecture to make training on laptops possible and improving training efficiency with the TF method. During the study, a total of three datasets are built and three models are trained for different objectives, namely:

- Detection of the concrete cracks;
- Detection of natural stone cracks;
- Differentiation between joints and cracks in natural stone.

This paper is organized as follows: in the next section, the design of proposed network architecture is presented; in Section 3, three datasets for each objective are build; section 4 demonstrates how the models are trained and how they benefit from the TF; section 5 shows the results of the study; section 6 makes conclusions of this article.

## Methodology and Implementation

### Network Architectures for Crack-like Features Detection

Each layer of the network has its own role in a deep learning network. The term convolution refers to an orderly mathematical procedure, in which two sources of information are intertwined and a new information is produced. The role of the convolution layer is a feature identifier (see Figure 1(a)). If features of input generally matched with filter, summation of multiplication will result in a large value in output (Fukushima, 1980). The Pooling (see Figure 1 (b)) layer can reduce the feature map size as the layers get deeper while at the same time keep the significant information. It helps to reduce the number of parameters and memory consumption in the network (Fukushima, 1980). The Rectified Linear Unit (ReLU) activation function (Glorot, 2011) is most commonly used in CNN based neural networks currently. With equation $R(x) = \max(0, x)$, the range of ReLU is $[0, \infty)$, which means that only a non-negative x-value yields and outputs. The uncomplicated and efficient mathematical form gives ReLU activation function layer a big advantage: It makes randomly initialized network very light, because of the characteristic of ReLU, approximately half of the neurons have 0 as output. This can cause several neurons to die and reduce parameters during training process. Fully Connected (FC) layer connects with high level features that extracted from convolutional layer with particular weights, and outputs the probabilities of different classes. As can be seen from Figure 2.14, in a FC layer, every neuron is connected with all the neurons in the previous layer. FC layers in CNN are identical to a fully connected multilayer perception structure. With suitable weight parameters, FC layers could create a stochastic likelihood representation.

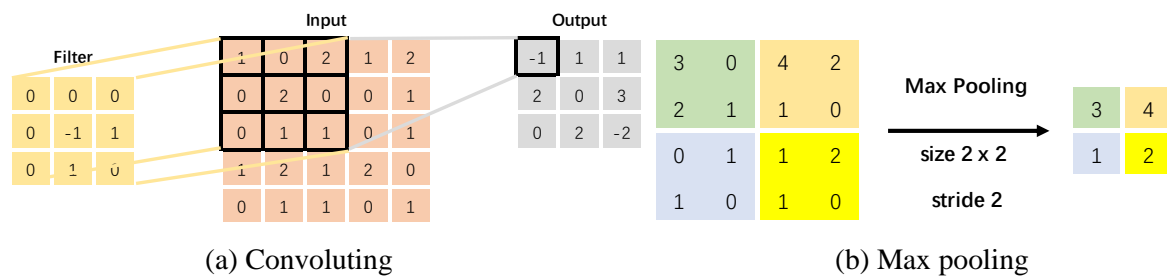

(a) Convoluting  (b) Max pooling

Figure 1: Convoluting and pooling (Glorot, 2011)

All popular network architectures are composed of above-mentioned basic layers. A comparison between most used architectures in Table 1 shows performance, depth and parameter number of those networks: On the one hand, the accuracy for image classification have been dramatically increased from 1998 to 2015. On the other hand, the architecture of network become deeper and more complex. In other words, the computer needs to process more than 60 million parameters to train a model.

| Name | LeNet-5 | AlexNet | VGG | GoogLeNet | ResNet |
|---|---|---|---|---|---|
| Year | 1998 | 2021 | 2014 | 2014 | 2015 |
| Top-5-Error | - | 15.30% | 7.30% | 6.67% | 3.57% |
| Data Augmentation | - | + | + | + | + |
| Number of Convolutional Layers | 3 | 5 | 16 | 21 | 151 |
| Layer Number | 7 | 8 | 19 | 22 | 152 |
| Parameter Number | 6.E+04 | 6.E+07 | 1.E+08 | 7.E+06 | 6.E+07 |

Table 1: Comparing between different CNN architecture (Russakovsky et al., 2015)

After the comparison, LeNet-5 is chosen to be basic architecture in this study for following reasons: First and most important, it meets the need of computational cost. Second, it has an acceptable accuracy. LeNet-5 is a classic CNN architecture proposed by Yann LeCun (1998). It was applied in banking to recognize handwritten numbers on checks. Because of the limited computing power at that time, grayscale images in 32 ×32 pixel is considered as inputs. LeNet-5 has 7 layers, 3 of them are convolutional layers. (see Figure 2).

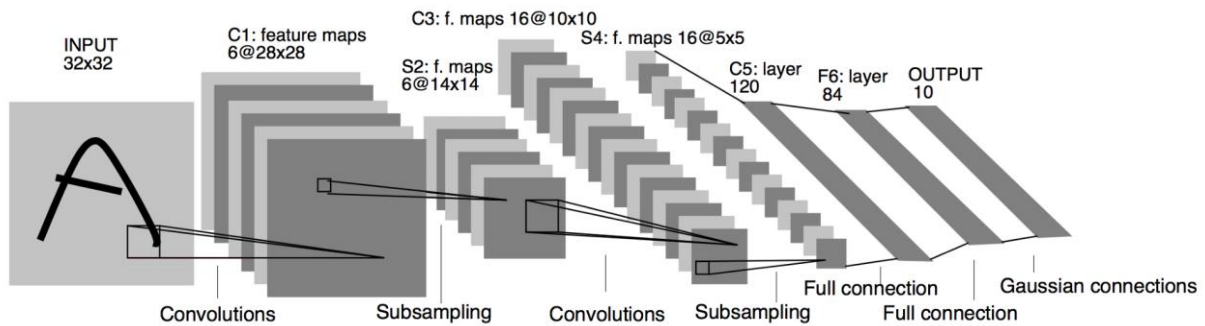

Figure 2: LeNet-5 architecture (Lecun, Bottou, Bengio and Haffner, 1998)

A number of modifications based on LeNet-5 need to be done to fit this architecture to the research goal: 1) Instead of one channel (black and white) images of original LeNet-5, the modified architecture takes three channels colour images as input. All inputs are re-sized into 228×228 pixels to avoid calculation errors, which caused by different image sizes in the dataset. 2) Instead of Sigmoid activation function of original network, which suffers from vanishing gradient problem, the modified architecture uses ReLU as activation function. 3) Max-Pooling and Local Response Normalization is used to keep features on the feature map. 4) The single 5x5 CONV layer is replaced by a stack of two 3x3 CONV layer to reduce parameters. A multi-FC layer set composed of FC1 and FC2 gives the network a stronger expression ability nonlinear to connect those extracted features from previous layers. 6) During the CNOV operation, SAME padding technique is used, which uses zeros to pad around the image to make sure the size of output and input are same. 7) On the end of the network architecture is a SoftMax layer to calculate the probabilities for each class. The modified network architecture is shown in Figure 3. Size and Parameters of each layer are shown in Table 2.

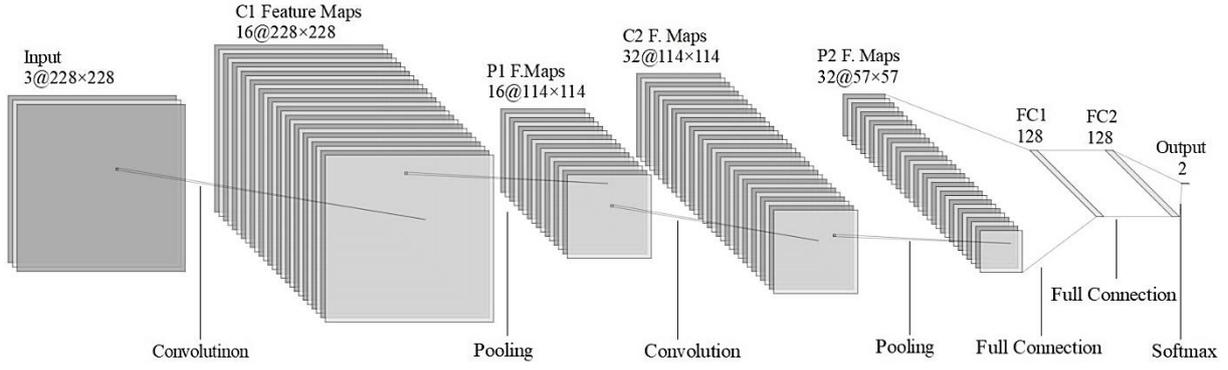

Figure 3:   Network architecture for crack-like feature detection

| Layer | Input Volume | | | Filter Size | | | Stride | | Output Volume | | | Parameters |
|---|---|---|---|---|---|---|---|---|---|---|---|---|
| | $D_i$ | $H_i$ | $W_i$ | K | $F_x$ | $F_y$ | $S_x$ | $S_y$ | $D_o$ | $H_o$ | $W_o$ | |
| Input | 3 | 228 | 228 | - | - | - | - | - | 3 | 228 | 228 | 0 |
| C1 | 3 | 228 | 228 | 16 | 3 | 3 | 1 | 1 | 16 | 228 | 228 | 448 |
| P1 | 16 | 228 | 228 | | 3 | 3 | 2 | 2 | 16 | 114 | 114 | 0 |
| C2 | 16 | 114 | 114 | 32 | 3 | 3 | 1 | 1 | 32 | 114 | 114 | 4640 |
| P2 | 32 | 114 | 114 | | 3 | 3 | 2 | 2 | 32 | 57 | 57 | 0 |
| FC1 | 32 | 57 | 57 | 1 | - | 128 | - | - | 1 | 128 | - | 13308032 |
| FC2 | 1 | 128 | - | 1 | - | 128 | - | - | 1 | 2 | | 258 |

Table 2:   Size and parameters of network architecture for crack-like feature detection

With the parameters trained after the transfer learning process for the target task, the size of SoftMax layer should also be changed (see Table 3). For example: to make the model able to different joints and crack in natural stone, the size of SoftMax layer is set to be 3 for different prediction results, namely cracks, joints and no defects.

| Training goals | Size of SoftMax layer |
|---|---|
| Detection of the concrete cracks | 2 |
| Detection of natural stone cracks | 2 |
| Differentiation between joints and cracks in natural stone | 3 |

Table 3:   Changes of SoftMax layer for different training goals

**Datasets**

The concrete cracks dataset (see Figure 4 (a)) contain a total of 40,000 images with 227×227 pixel resolutions (Zhang, 2016). The whole dataset is evenly divided into two groups as positive crack and negative crack images for classification. Data augmentation like random rotation or flipping is not used in this dataset. This dataset is divided into a training set with 39600 images and a test set with 2400 images.

The natural stone cracks dataset (see Figure 4 (b)) contain 150 small images with different Resolutions. The whole dataset is divided into a test set of 10 images and a train set of 140 images. In the training set, 70 images are labelled with crack and other 70 images are labelled with negative. Different types of cracks are contained in the dataset, such as hair crack and splitting.

The cracks and joints of natural stones dataset (see Figure 4 (c)) is made of natural stone contains 150 images with different pixel resolutions. All images related to natural stone cracks are cut from RGB images with a resolution of 2560×1920. The whole dataset is divided into a test set of 10 images and a train set of 140 images. The training set consists of 70 pictures with natural stone crack and 70 pictures with joints. In the test sets, 5 images are natural stone with cracks and others images are about joints.

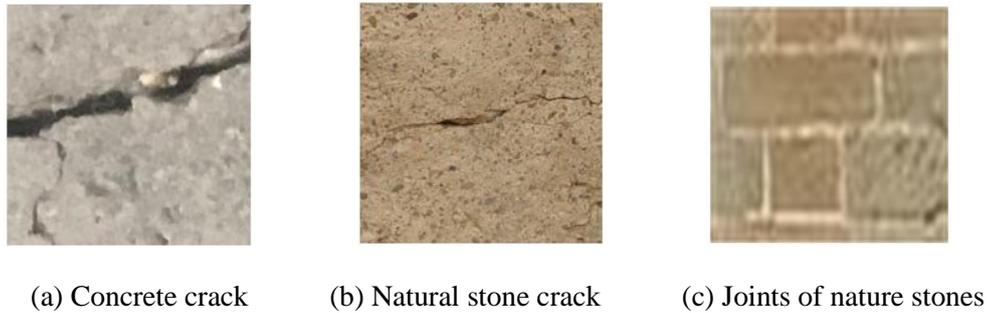

(a) Concrete crack     (b) Natural stone crack     (c) Joints of nature stones

Figure 4: Examples of cracks and joints in datasets

**Training Process and Results**

An overview of training process can be seen from Figure 5: Step 1: Feed the images and labels into the network. Step 2: Iterate over each example in the training dataset within one step by grabbing its features and label. Step 3: Compare the prediction of inputs with the real label. Measure the SoftMax value of the prediction and use that to calculate the model's loss and gradients. Step 4: Update the model's variables with Adam optimizer. Step 5: Repeat for each epoch. The loss value ls calculated with $Loss = logS_j$. The SoftMax function can be described with $S_j = e^{a_j}/\sigma_{k=1}^{T} e^{a_k}$, where $S_j$ is the SoftMax value of element j, $a_j$ is the original value of the element, and $T$ is the number of elements in the vector.

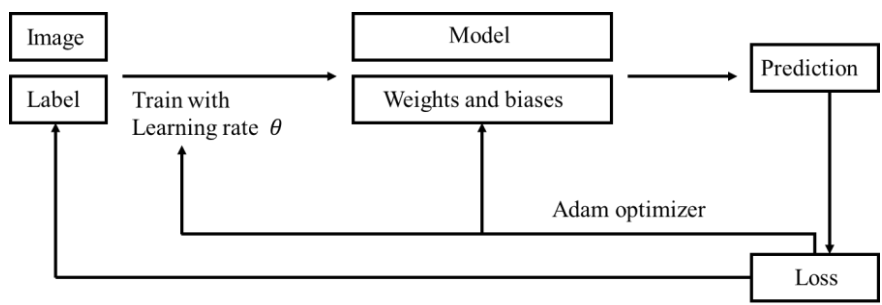

Figure 5: Training process overview

The Training process of the **concrete cracks detection model** took 3038 seconds. Figure 6 summarizes the loss and accuracy changes during training process of this model. As can be seen from Figure 6 (a), those lines in purple and orange show the fluctuation of training loss and testing loss separately. They both show sharp fluctuation before 10,800 training steps and trend to become steady after 11,100 training steps. Until 15,000 training steps, the accuracy of testing gets stable around 100%. Taking the model with 14,199 training steps for example, when the test loss is near to 0, the test accuracy is near to 1. We hence keep the model with 14,199 training steps for validation and the TF.

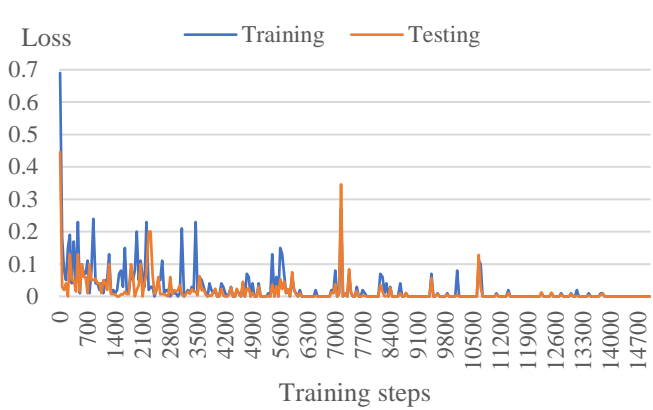
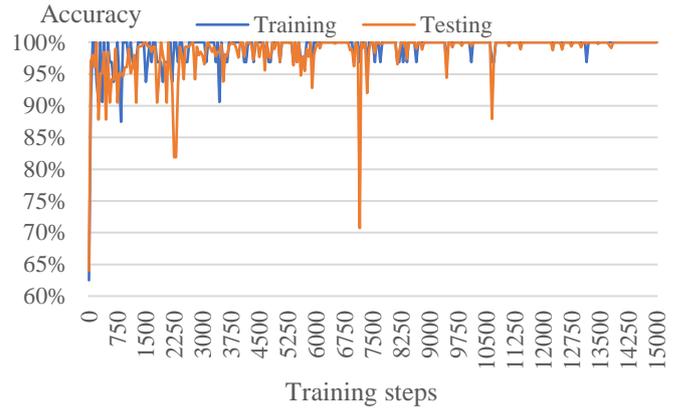

(a) Loss

(b) Accuracy

Figure 6: Loss and accuracy changes of concrete crack detection model

As can be seen in Figure 7(a) for natural stone cracks detection model, the red line stands for the accuracy using the TF, and the light purple line stands for the accuracy without using the TF. After first 50 training steps, training accuracy of the TF is 87.5%, which is higher than the one without (53.12%). Additionally, it took 36 seconds for training with TF to get a training accuracy over 97%, while it took 62 seconds for the training without TF. Similarly for Cracks and joints detection model, Figure 7(b) shows that TF needs less time (28 seconds) compared to normal training (44 seconds) to get an accuracy over 97%.

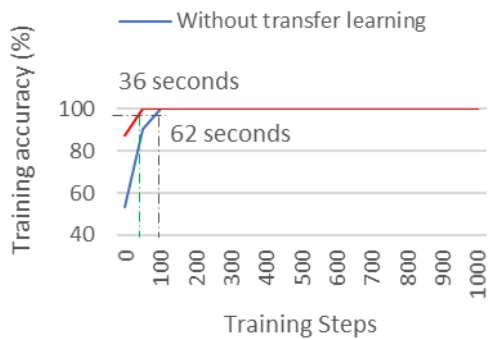
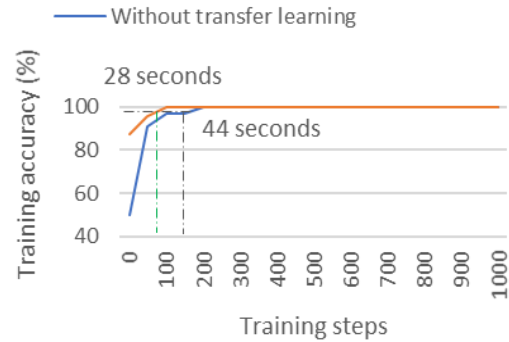

(a) Natural stone cracks detection model

(b) Cracks and joints detection model

Figure 7: Comparison between training with and without the TF

The training process of **the natural stone cracks detection model** took 367 seconds. Figure 8 provides data regarding loss and accuracy changes while the model was training. Both training accuracies and testing accuracies keep steady after 150 train steps, where training accuracy is around 100 % and testing accuracy is around 90 %.

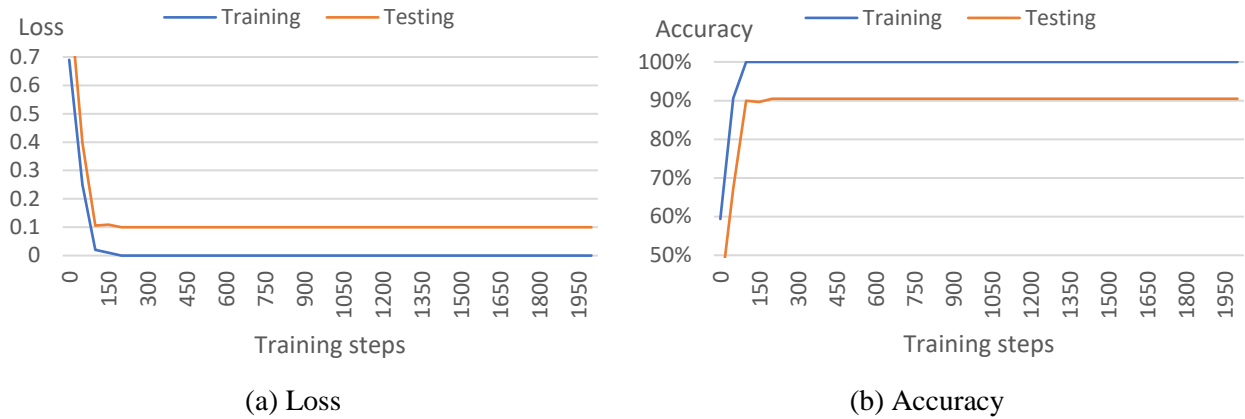

(a) Loss  (b) Accuracy

Figure 8 Loss and accuracy changes of natural stone crack detection model

The training process of the **cracks and joints of natural stones detection model** took 84 seconds. As can be seen in Figure 9, training accuracy and testing accuracy increase to around 100 % and around 82 separately after 400 steps.

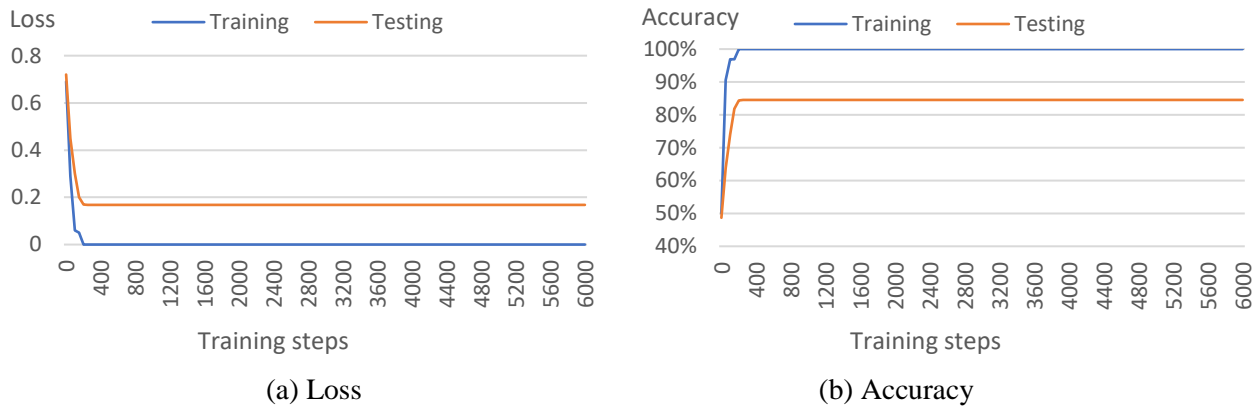

(a) Loss  (b) Accuracy

Figure 9: Loss and accuracy changes of cracks and joints detection model

**Results**

After loading the **concrete cracks detection model** with 141999 training steps, Figure 11(a) and 10 (b) shows the prediction results of a concrete with 0.506732 probability of having no crack and with 1.000000 probability of having a crack separately. After loading the **natural stone cracks detection model** with 1999 training steps, Figure 11(c) and 10 (d) shows the prediction results of a natural stone with 0.999998 and 0.951548 probability to have a crack separately.

It should be pointed out that the probability value indicates the confidence degree of the computer on the prediction result. It is calculated with $probability = e^{-Loss}$.

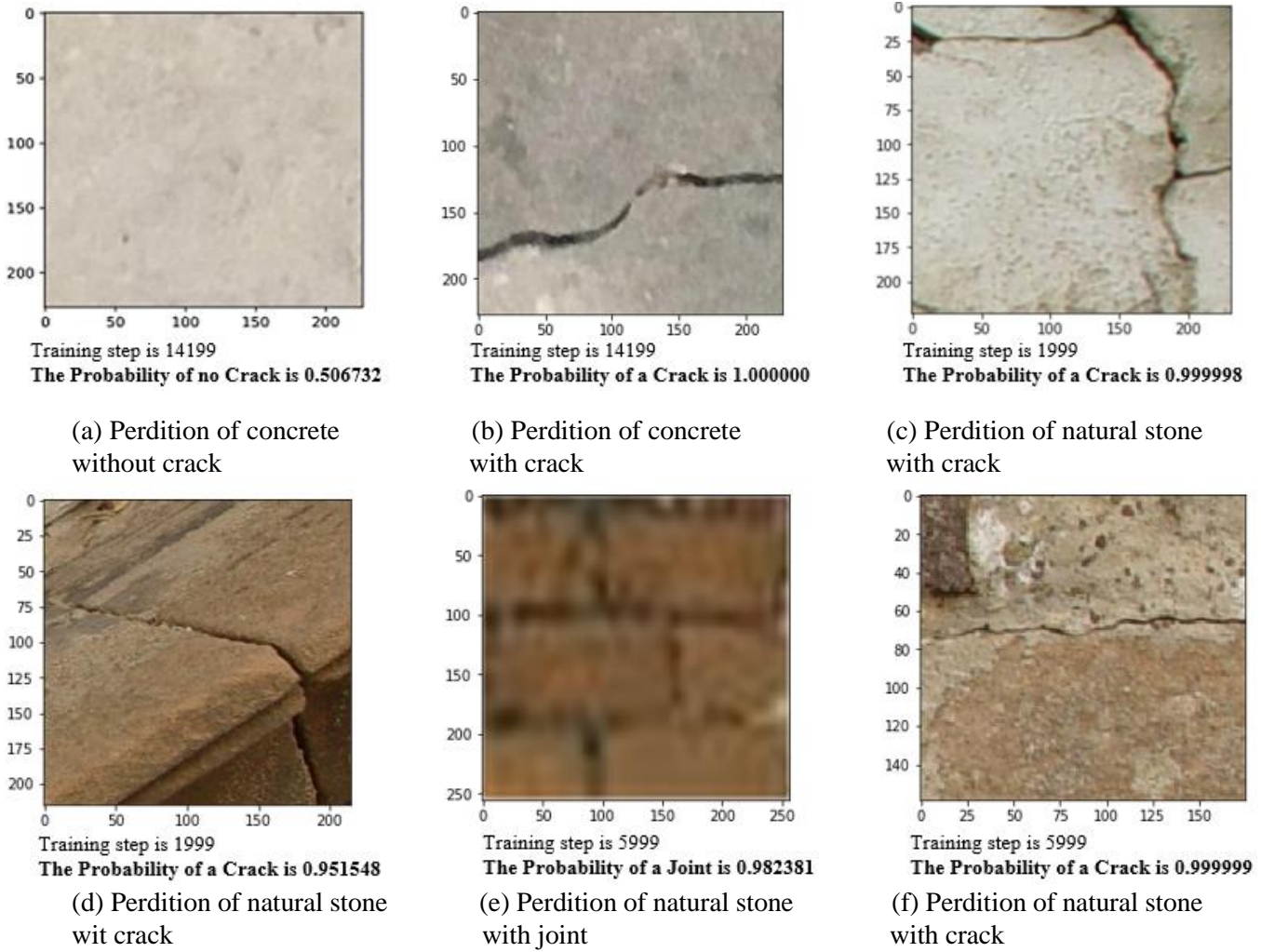

| (a) Perdition of concrete without crack | (b) Perdition of concrete with crack | (c) Perdition of natural stone with crack |
| --- | --- | --- |
| (d) Perdition of natural stone wit crack | (e) Perdition of natural stone with joint | (f) Perdition of natural stone with crack |

Figure 11: Utilization of the models

**Conclusion and Discussion**

The whole study consists mainly of four parts. A lightweight CNN architecture is built. A CNN model is then trained so that it can detect whether there are concrete cracks in images. After that, the TF method is used in the training process to train a natural stone detection model with few training data. Consequently, the training steps and training time to get the comparable result are significantly reduced. Since there are not only cracks in a natural stone façade, a further advanced CNN, to promote the usage scenarios of this CNN in renovation works, is trained, so that the computer is able to distinguish, if there are cracks, joints, or nothing special in the input image.

From the comparative studies, which also used CNN to detect concrete cracks, this proposed light weighted CNN architecture shows a possibility to training models on laptops compared to the Cha (2017). Training time of all three models in this work are within one hour. Also, our results suggest that the main advantages of transfer learning are the potential of saving training time as well as solve the problem of insufficient training data (see Figure 7). As shown in results, our work is demonstrated as an effective tool for the industry use.

Although there are discoveries revealed by this study, there are also limitations. The size of data set with cracks is quite small, which makes the trained model not robust enough because not all the features from diverse cracks are including in the dataset. Thanks to the relative lightweight architecture of our model, we encourage future research to implement our model on mobile applications, such as smart phone, to make the renovation process more efficient and smoother.

**References**

LeCun, Y., Bottou, L., Bengio, Y. and Haffner, P., 1998. Gradient-based learning applied to document recognition. Proceedings of the IEEE, 86(11), pp.2278-2324.

Krizhevsky, A., Sutskever, I. and Hinton, G.E., 2012. Imagenet classification with deep convolutional neural networks. Advances in neural information processing systems, 25, pp.1097-1105.

Simonyan, K. and Zisserman, A., 2014. Very deep convolutional networks for large-scale image recognition. arXiv preprint arXiv:1409.1556.

He, K., Zhang, X., Ren, S. and Sun, J., 2016. Deep residual learning for image recognition. In Proceedings of the IEEE conference on computer vision and pattern recognition (pp. 770-778).

Szegedy, C., Liu, W., Jia, Y., Sermanet, P., Reed, S., Anguelov, D., Erhan, D., Vanhoucke, V. and Rabinovich, A., 2015. Going deeper with convolutions. In Proceedings of the IEEE conference on computer vision and pattern recognition (pp. 1-9).

Cha, Y.J., Choi, W. and Büyüköztürk, O., 2017. Deep learning-based crack damage detection using convolutional neural networks. Computer-Aided Civil and Infrastructure Engineering, 32(5), pp.361-378.

Kang, D. and Cha, Y., 2018. Autonomous UAVs for Structural Health Monitoring Using Deep Learning and an Ultrasonic Beacon System with Geo-Tagging. Computer-Aided Civil and Infrastructure Engineering, 33(10), pp.885-902.

Fukushima, K., 1980. Neocognitron: A self-organizing neural network model for a mechanism of pattern recognition unaffected by shift in position. Biological Cybernetics, 36(4), pp.193-202.

Glorot, X., Bordes, A. and Bengio, Y., 2011, June. Deep sparse rectifier neural networks. In Proceedings of the fourteenth international conference on artificial intelligence and statistics (pp. 315-323). JMLR Workshop and Conference Proceedings.Russakovsky, O., Deng, J., Su, H., Krause, J., Satheesh, S., Ma, S., Huang, Z., Karpathy, A., Khosla, A., Bernstein, M., Berg, A. and Fei-Fei, L., 2015. ImageNet Large Scale Visual Recognition Challenge. International Journal of Computer Vision, 115(3), pp.211-252.

Zhang, L., Yang, F., Zhang, Y.D. and Zhu, Y.J., 2016, September. Road crack detection using deep convolutional neural network. In 2016 IEEE international conference on image processing (ICIP) (pp. 3708-3712). IEEE.

Yosinski, J., Clune, J., Bengio, Y. and Lipson, H., 2014. How transferable are features in deep neural networks?. arXiv preprint arXiv:1411.1792.

Xiang, Y., Tang, T., Su, T., Brach, C., Liu, L., Mao, S. and Geimer, M., 2020. Fast crdnn: Towards on site training of mobile construction machines. arXiv preprint arXiv:2006.03169.

Xiang, Y., Wang, H., Su, T., Li, R., Brach, C., Mao, S.S. and Geimer, M., 2020. Kit moma: A mobile machines dataset. arXiv preprint arXiv:2007.04198.

ASHRAE, (2005). Handbook of Fundamentals. Atlanta: American Society of Heating, Refrigerating and Air Conditioning Engineers.

Abdel-Qader, I., Abudayyeh, O. and Kelly, M.E., 2003. Analysis of edge-detection techniques for crack identification in bridges. Journal of Computing in Civil Engineering, 17(4), pp.255-263.

Yeum, C.M. and Dyke, S.J., 2015. Vision-based automated crack detection for bridge inspection. Computer-Aided Civil and Infrastructure Engineering, 30(10), pp.759-770.

Zhou, B., Khosla, A., Lapedriza, A., Oliva, A. and Torralba, A., 2014. Object detectors emerge in deep scene cnns. arXiv preprint arXiv:1412.6856.

Szegedy, C., Vanhoucke, V., Ioffe, S., Shlens, J. and Wojna, Z., 2016. Rethinking the inception architecture for computer vision. In Proceedings of the IEEE conference on computer vision and pattern recognition (pp. 2818-2826).

Smith, L.N., 2017, March. Cyclical learning rates for training neural networks. In 2017 IEEE winter conference on applications of computer vision (WACV) (pp. 464-472). IEEE.

Khan, S., Rahmani, H., Shah, S.A.A. and Bennamoun, M., 2018. A guide to convolutional neural networks for computer vision. Synthesis Lectures on Computer Vision, 8(1), pp.1-207.

Kang, D. and Cha, Y.J., 2018. Autonomous UAVs for structural health monitoring using deep learning and an ultrasonic beacon system with geo-tagging. Computer-Aided Civil and Infrastructure Engineering, 33(10), pp.885-902.

Neto, N. and De Brito, J., 2011. Inspection and defect diagnosis system for natural stone cladding. Journal of Materials in Civil Engineering, 23(10), pp.1433-1443.

Cs.toronto.edu. 2021. CSC321 Winter 2018.

Satyen, R., 2019. [online] Available at: <https://github.com/satyenrajpal/Concrete-Crack-Detection> [Accessed 17 May 2020].

Katex NDT Equipment. 2021. UV Magnetic Yoke Flaw Detector - AJE-220 | Katex Ltd. [online] Available at: <https://www.katex.co.uk/product/aje-220-ac-uv-magnetic-yoke-flaw-detector/> [Accessed 19 May 2021].

eBay. 2021. Olympus Panametrics Omniscan MX 32 : 128 Ultraschall Phased Feld Pa Flaw | eBay. [online] Available at: <https://www.ebay.de/itm/393249231234?mkevt=1&mkcid=1&mkrid=707-53477-19255-0&campid=5338805019&toolid=20006&customid=1250078041622-g_Cj0KCQjw7pKFBhDUARIsAFUoMDZU_S37KnAtbmEdSh96x15OBk9A7oD-IPHHHrAL7bSiE2EOZeD-ycAaArPOEALw_wcB> [Accessed 19 May 2021].